# Influence of vacuum annealing on interface properties of SiC (0001) MOS structures

Koji Ito[1]*, Takuma Kobayashi[1], and Tsunenobu Kimoto[1]

[1]*Department of Electronic Science and Engineering, Kyoto University, Nishikyo, Kyoto 615-8510, Japan*

*E-mail: ito@semicon.kuee.kyoto-u.ac.jp

We investigated the influence of vacuum annealing on interface properties of silicon carbide (SiC) metal-oxide-semiconductor (MOS) structures. For as-oxidized and nitric oxide (NO)-annealed samples, the interface state density ($D_{it}$) near the conduction band edge ($E_C$) of SiC did not increase by subsequent vacuum annealing. For phosphoryl chloride ($POCl_3$)-annealed samples, in contrast, $D_{it}$ at $E_C - 0.2$ eV increased from $1.3 \times 10^{10}$ to $2.2 \times 10^{12}$ cm$^{-2}$eV$^{-1}$ by the vacuum annealing, and the channel mobility of MOS field effect transistors (MOSFETs) decreased from 109 to 44 cm$^2$V$^{-1}$s$^{-1}$. Mechanism of the observed increase in $D_{it}$ was discussed based on the results of secondary ion mass spectrometry measurement.

Silicon carbide (SiC) owns superior material properties for power device applications, such as wide bandgap and high critical electric field.[1,2] Thus, SiC metal-oxide-semiconductor field effect transistors (MOSFETs) are expected as power switching devices with low-loss and high-speed operation. In spite of their great potential, SiC MOSFETs have suffered from the low channel mobility due to the extremely high interface state density ($D_{it}$, ~ $10^{13}$ cm$^{-2}$eV$^{-1}$) of silicon dioxide (SiO$_2$)/SiC systems.[3–8]

Although the physical origin of interface states is unclear, it was found that interface nitridation (annealing in nitric oxide (NO)[9–13] or nitrous oxide (N$_2$O)[11,14,15]) reduces the $D_{it}$ to about $10^{12}$ cm$^{-2}$eV$^{-1}$ and improves the channel mobility ($\mu_{ch}$) of the MOSFETs up to ~ 30 cm$^2$V$^{-1}$s$^{-1}$. Nitrogen atoms segregate in the vicinity of the interface by the nitridation[14] and there exists certain relationship between the nitrogen concentration at the interface and the $D_{it}$.[16,17] Phosphorus treatment (annealing in a gas mixture of phosphoryl chloride (POCl$_3$), oxygen (O$_2$), and nitrogen (N$_2$)) also reduces the $D_{it}$ to about $10^{11}$ cm$^{-2}$eV$^{-1}$ and improves $\mu_{ch}$ up to ~ 89 cm$^2$V$^{-1}$s$^{-1}$.[18,19] By this treatment, phosphorus atoms almost uniformly distribute in the oxide with a density of typically $10^{21}$ cm$^{-3}$, leading to formation of phosphosilicate glass (PSG). Although the interface properties are sensitively affected by the introduction of foreign atoms as described above, the mechanism of the defect passivation has not been well understood yet.

In actual power MOSFETs, since it is necessary to reduce the contact resistivity ($\rho_c$) to about $10^{-6}$ $\Omega cm^2$, additional annealing should be performed after the formation of a gate oxide film; A typical condition is two-step annealing at 600°C for 3 min and 1000°C for 2 min.[20] Although the annealing may affect interface properties, there are very few reports focusing on the influence of this annealing. In this study, we thus investigated the effect of the annealing on the MOS interface properties as well as electrical characteristics of MOSFETs.

MOS capacitors and MOSFETs were fabricated on 4° off-axis *n*-type (donor density: $N_D$ = 5×10$^{15}$ cm$^{-3}$) and 8° off-axis *p*-type (acceptor density: $N_A$ = 3×10$^{15}$ cm$^{-3}$) 4H-SiC (0001), respectively. The gate oxides were formed by dry oxidation at 1300°C for 30 min (As-Ox.), by oxidation with subsequent NO (10% diluted in $N_2$) annealing at 1250°C for 70 min (Ox.+NO), or by oxidation with subsequent POCl$_3$ (a gas mixture of POCl$_3$, $O_2$, and $N_2$) annealing at 1000°C for 10 min. After the POCl$_3$ annealing, $N_2$ annealing at 1000°C for 30 min was also performed. The resulting oxide thicknesses of as-oxidized, NO-annealed and POCl$_3$-annealed samples were in the range of 42-46, 41-45 and 55-57 nm, respectively. To investigate the effect of vacuum annealing, the SiO$_2$/SiC samples were subsequently annealed at 600°C for 3 min and 1000°C for 2 min in vacuum. Circular aluminum (Al) gate electrodes with a diameter of about 300 - 500 μm were formed for MOS capacitors. The gate metal, channel length, and channel width of MOSFETs were Al, 50 - 770 μm, and 100 - 200 μm, respectively. All of the measurements were conducted at room temperature.

Figure 1 shows the energy distributions of $D_{it}$ obtained by a high (1 MHz)-low[21–23] or $C$-$\psi_S$ method[24] for (a) as-oxidized, (b) NO-annealed, and (c) POCl$_3$-annealed MOS structures with and without the two-step vacuum annealing. We see that the subsequent vacuum annealing has a limited influence on the $D_{it}$ distributions in the case of as-oxidized and NO-annealed samples. On the other hand, the $D_{it}$ value at $E_C - 0.2$ eV estimated by the high (1 MHz)-low method increased from $1.3 \times 10^{10}$ to $2.2 \times 10^{12}$ cm$^{-2}$eV$^{-1}$ by the vacuum annealing in the case of the POCl$_3$-annealed sample (Fig. 1 (c)).

Note that the $C$-$\psi_S$ method can detect very fast states responding to high frequencies over 1 MHz,[24] which cannot be detected by the conventional high (1 MHz)-low method. In Figs. 1 (a), (b), and (c), the $D_{it}$ values extracted by the $C$-$\psi_S$ method are always higher than those estimated by the high (1 MHz)-low method, due to the existence of fast interface states. For the NO-annealed sample (Fig. 1 (b)), the difference between the $D_{it}$ values obtained by high (1 MHz)-low and $C$-$\psi_S$ methods is particularly large (e.g., $1.9 \times 10^{12}$ cm$^{-2}$eV$^{-1}$ @$E_C - 0.2$ eV), due to the generation of very fast states by the nitridation treatment.[16] For as-oxidized and NO-annealed samples (Figs. 1 (a) and (b)), neither the density of slow states nor the fast states was almost affected by the vacuum annealing. For the POCl$_3$-annealed samples (Fig. 1 (c)), on the other hand, both the slow and fast states significantly increased by the annealing. For instance, the $D_{it}$ at $E_C - 0.2$ eV estimated by the high (1 MHz)-low method increased from $1.3\times10^{10}$ to $2.2\times10^{12}$ cm$^{-2}$eV$^{-1}$ and that by the $C$-$\psi_\mathrm{S}$ method increased from $6.1\times10^{10}$ to $3.7\times10^{12}$ cm$^{-2}$eV$^{-1}$ by the vacuum annealing.

Figure 2 depicts the gate voltage dependence of field-effect mobility for as-oxidized, NO-annealed, and POCl$_3$-annealed MOSFETs, with and without the two-step vacuum annealing. Degradation of the field-effect mobility by vacuum annealing was remarkable for the POCl$_3$-annealed sample, being in consistent with the results of $D_\mathrm{it}$ estimation (Fig. 1 (c)); The peak value of the mobility decreased from 109 to 44 cm$^2$V$^{-1}$s$^{-1}$ by the vacuum annealing. In contrast, the mobilities of as-oxidized and NO-annealed MOSFETs samples were not so affected by the vacuum annealing as the POCl$_3$-annealed sample, which is also consistent with the $D_\mathrm{it}$ distribution (Figs. 1 (a) and (b)).

Figure 3 shows depth profiles of nitrogen and phosphorus concentrations in NO- and POCl$_3$-annealed SiO$_2$/SiC samples, respectively, with and without two-step vacuum annealing, measured by using secondary ion mass spectrometry (SIMS). For the NO-annealed samples, small increase in the $D_\mathrm{it}$ (Fig. 1 (b)) or small deterioration in the $\mu_\mathrm{ch}$ (Fig. 2) occurred by the vacuum annealing, which does not contradict to the fact that little change was also detected in the nitrogen profile with the annealing. Majority of the nitrogen atoms remain at the interface to passivate the interface defects even after the vacuum annealing, leading to the suppression of increase in the $D_\mathrm{it}$. For the POCl$_3$- annealed sample, although remarkable increase in the $D_{it}$ (Fig. 1 (c)) and deterioration in the $\mu_{ch}$ (Fig. 2) were observed after the vacuum annealing, the phosphorus profile was not substantially affected by the annealing.

We here briefly discuss the possible origin of the observed increase in the $D_{it}$ by the vacuum annealing in the POCl$_3$-annealed sample. It is reported that PSG formed by the POCl$_3$ annealing contains −O$_3$PO configurations with dangling oxygen atoms in its network, and that such oxygen atoms may easily desorb in the O-poor ambient, resulting in the generation of −O$_3$P configurations.[25] Thus, during the vacuum annealing, the dangling oxygen atoms may desorb from the −O$_3$PO configurations and subsequently attack the SiO$_2$/SiC interface to proceed the oxidation of SiC, which leads to the creation of interface defects such as carbon-related defects.[2,26–33] However, as the oxide thickness measured by spectroscopic ellipsometry did not change with the vacuum annealing, we speculate that the oxidation of SiC during the vacuum annealing only proceeds for a few atomic layers or so.

In summary, we investigated the influence of vacuum annealing on interface properties of SiO$_2$/SiC systems. As a result, for as-oxidized and NO-annealed samples, $D_{it}$ values near $E_C$ were almost unchanged by subsequent vacuum annealing. On the other hand, for POCl$_3$-annealed samples, $D_{it}$ at $E_C - 0.2$ eV increased from $1.3 \times 10^{10}$ to $2.2 \times 10^{12}$ cm$^{-2}$eV$^{-1}$ by the annealing, and the channel mobility of MOSFETs decreased from 109 to 44

cm$^2$V$^{-1}$s$^{-1}$. Depth profiles of nitrogen and phosphorus concentrations in NO- and POCl$_3$-annealed samples, respectively, were not affected by vacuum annealing. Based on this result, we suggested that the dangling oxygen atoms of −O$_3$PO configurations in PSG may desorb and attack the SiO$_2$/SiC interface during the vacuum annealing, leading to the oxidation of SiC and increase in the $D_{it}$.

This work was supported in part by the Super Cluster Program and Open Innovation Platform with Enterprises, Research Institute and Academia (OPERA) Program from the Japan Science and Technology Agency.

## Figure Captions

**Fig. 1.** Energy distributions of interface state density for as-oxidized, NO-annealed, and POCl$_3$-annealed SiC MOS capacitors with and without two-step vacuum annealing at 600°C for 3 min and 1000°C for 2 min extracted by a high (1 MHz)-low or $C$-$\psi_S$ method.

**Fig. 2.** Gate voltage dependence of field-effect mobility for as-oxidized, NO-annealed, and POCl$_3$-annealed SiC MOSFETs with and without two-step vacuum annealing at 600°C for 3 min and 1000°C for 2 min.

**Fig. 3.** Depth profiles of nitrogen and phosphorus concentrations in NO- and POCl$_3$-annealed SiO$_2$/SiC samples, respectively, with and without two-step vacuum annealing at 600°C for 3 min and 1000 for 2 min. The profiles were measured by using secondary ion mass spectrometry (SIMS). The abscissa was shifted so that the position of the SiO$_2$/SiC interface becomes same for the four data.

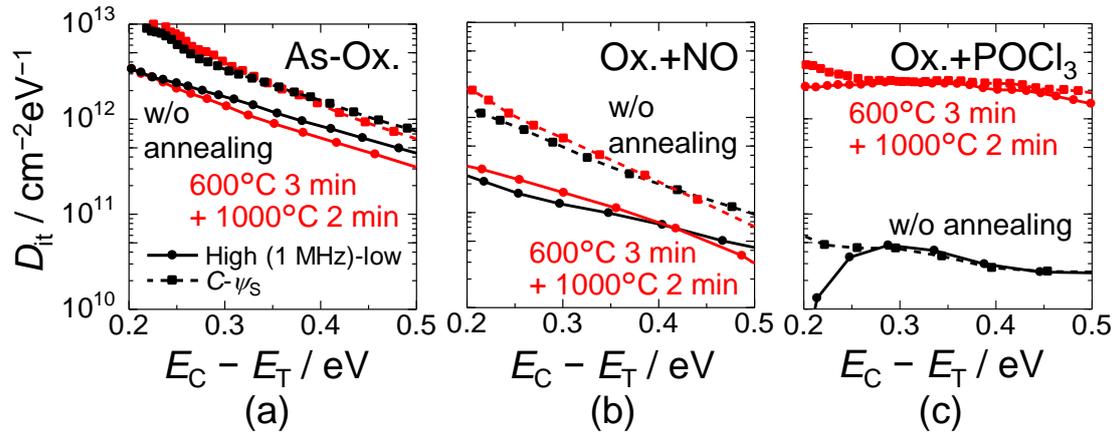

Fig. 1.

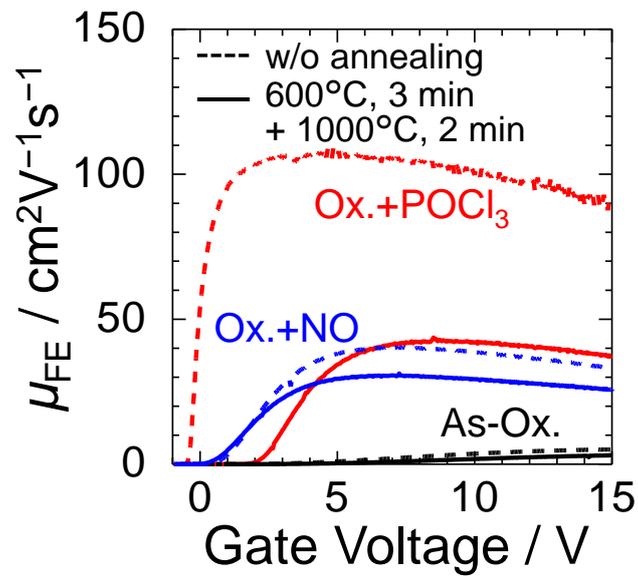

Fig. 2.

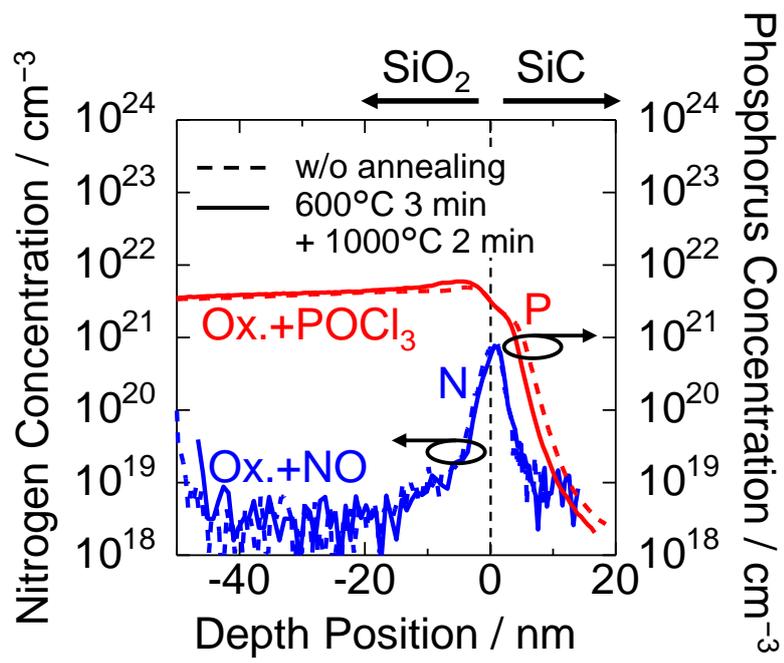

Fig. 3.